%% file: aaai23.tex
\documentclass[letterpaper]{article} 
\usepackage{aaai23}  
\usepackage{times}  
\usepackage{helvet}  
\usepackage{courier}  
\usepackage[hyphens]{url}  
\usepackage{graphicx} 
\usepackage{natbib}  
\usepackage{caption} 
\usepackage{subfigure}
\usepackage{multirow}
\usepackage{amssymb}
\usepackage{soul}
\usepackage{url}
\usepackage[utf8]{inputenc}
\usepackage{amsthm}

\usepackage{newfloat}
\usepackage{listings}
\DeclareCaptionStyle{ruled}{labelfont=normalfont,labelsep=colon,strut=off} 
\usepackage{amsmath}
\usepackage{booktabs}
\usepackage{algorithm}
\usepackage{algorithmic}

\title{VarietySound: Timbre-Controllable Video to Sound Generation via Unsupervised Information Disentanglement}

\author{Chenye Cui\thanks{Equal contribution.}, Yi Ren\footnotemark[1], Jinglin Liu\footnotemark[1], Rongjie Huang\footnotemark[1], Zhou Zhao\thanks{Corresponding author}}
\affiliations{
Zhejiang University\\
\{chenyecui,rayeren,jinglinliu,rongjiehuang,zhaozhou\}@zju.edu.cn
}

\usepackage{bibentry}

\begin{document}

\maketitle

\input{Sections/0_abstract.tex}
\input{Sections/1_introduction.tex}
\input{Sections/2_related_works.tex}
\input{Sections/3_method.tex}

\input{Sections/5_experiments.tex}
\input{Sections/6_conclusion.tex}

\bibliography{aaai23}

\input{Sections/7_appendix.tex}

\end{document}

%% file: Sections/0_abstract.tex
\begin{abstract}
Video to sound generation aims to generate realistic and natural sound given a video input.
However, previous video-to-sound generation methods can only generate a random or average timbre without any controls or specializations of the generated sound timbre, leading to the problem that people cannot obtain the desired timbre under these methods sometimes. 
In this paper, we pose the task of generating sound with a specific timbre given a video input and a reference audio sample.
To solve this task, we disentangle each target sound audio into three components: temporal information, acoustic information, and background information.
We first use three encoders to encode these components respectively:
1) a temporal encoder to encode temporal information, which is fed with video frames since the input video shares the same temporal information as the original audio;
2) an acoustic encoder to encode timbre information, which takes the original audio as input and discards its temporal information by a temporal-corrupting operation;
and 3) a background encoder to encode the residual or background sound, which uses the background part of the original audio as input.
Then we use a decoder to reconstruct the audio given these disentangled representations encoded by three encoders.
To make the generated result achieve better quality and temporal alignment, we also adopt a mel discriminator and a temporal discriminator for the adversarial training.
In inference, we feed the video, the reference audio and the silent audio into temporal, acoustic and background encoders and then generate the audio which is synchronized with the events in the video and has the same acoustic characteristics as the reference audio with no background noise.
Our experimental results on the VAS dataset demonstrate that our method can generate high-quality audio samples with good synchronization with events in video and high timbre similarity with the reference audio.
Our demos have been published at \url{https://conferencedemos.github.io/Anonymous9517}.
\end{abstract}

%% file: Sections/1_introduction.tex
\section{Introduction}
\begin{figure}[htp]
\begin{center}
\includegraphics[width=0.9\linewidth]{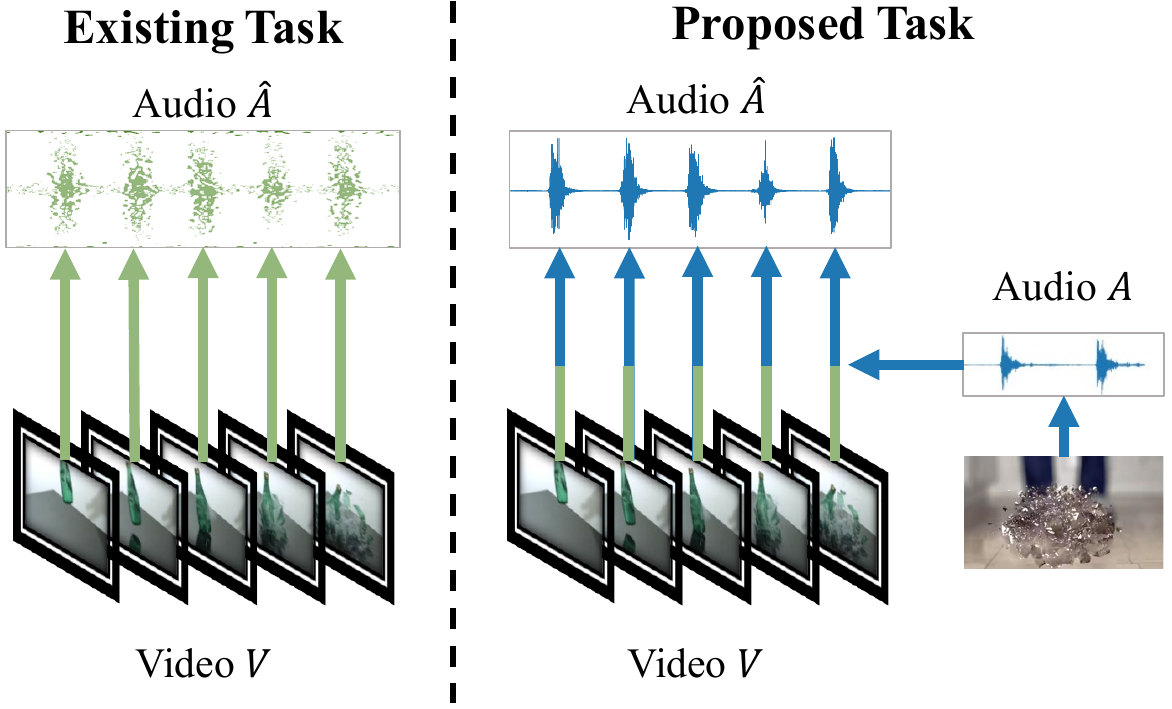}
\end{center}
   \caption{Timbre Controllable Video to Sound Generation}
\label{fig:task}
\vspace{-4mm}
\end{figure}
More and more people are devoting themselves to video creation, with the boom of microfilm and the rise of the short video market recently.
However, many people do not produce studio-level video works due to the high cost of post-production, which has become a critical issue affecting the development of the personal video production industry.
As machine learning continues to evolve in the video and audio fields~\cite{huang2022prodiff,huang2022fastdiff,cui2021emovie,lam2021bilateral,wang2022metarangeseg,zhang2022federated,zhang2021practical}, more and more AI-based software is being developed to help people create after-effects generation, background music production, video dubbing, and other post-production.
Video sound effects are an exceptionally critical part of video post-production, and a fantastic sound effect can bring a better video viewing experience to the audience.
The traditional methods of manual sound effects production are usually expensive and time-consuming or produce sounds that are often unnatural, therefore are not suitable for individual producers or do not provide an excellent audio-visual experience.

Several video-to-audio generation methods \cite{owens2016visually,zhou2018visual,chen2018visually,chen2020generating,ghose2021foleygan} have been proposed to solve the problem of dubbing for silent videos.
Nevertheless, they share a common problem: all of their acoustic information comes from the model's prediction and cannot control the timbre of the generated audio.
To match this problem, we defined a task called Timbre Controllable Video to Sound Generation (TCVSG), whose target is to allow users to generate realistic sound effects with their desired timbre for silent videos.
As illustrated in Figure \ref{fig:task}, we have a video clip \boldmath{V} of an object breaking for movie production, but the naturally recorded sounds are not impressive enough. 
So with this task, we can use an additional audio \boldmath{A} with a more remarkable sound of breaking to generate an audio track for \boldmath{V}.
The generated audio \boldmath{\^{A}} will be time-aligned with \boldmath{V}, but has the same kind of sound as \boldmath{A} which will make the video more exciting.
As far as we know, we are the first to propose this task.

To accomplish this task, in this work, we propose VarietySound, a generative video-to-sound model for generating high-quality audio which is synchronized with the events in video and has the same acoustic characteristics as the reference audio.
As is demonstrated in SPEECHSPLIT \cite{qian2020unsupervised}, audio can be disentangled into multiple independent pieces of information and modeled separately for the audio reconstruction or timbre transfer.
Some other previous works \cite{aytar2016soundnet,owens2016ambient,takahashi2017aenet}  have shown a strong correlation between audio and video in the temporal order.
Therefore, we use a method based on the information disentanglement, which first disentangles the information of an audio into three parts and then recombines these information to reconstruct the audio.
Specifically, in the training stage, we use three different encoders to modeling the information above:
\begin{itemize}
    \item A temporal encoder to modeling the temporal information which is provided through the video features and employs a bottleneck to the output to avoid leaking other content information;
    \item An acoustic encoder to modeling the timbre information which is provided through the original audio and conducts a temporal-corrupting operation to discard the temporal information;
    \item A background encoder to modeling the background information, which is also provided through the original audio but applies a mask method to make sure only the background is utilized;
\end{itemize}
and use a decoder that accepts the encoders' outputs as input to recombine these information for generating the final audio.

In order to get a quality close to real audio and a better temporal alignment to the video, we use two discriminators for adversarial training:
\begin{itemize}
    \item A multi-window mel discriminator to discriminate the quality of the generated mel-spectrogram; 
    \item A time-domain alignment discriminator to discriminate whether the audio has temporal alignment with the video.
\end{itemize}

In the inference stage, we feed the silent video, the reference audio and the silent audio into temporal, acoustic and background encoders and then generate time-aligned, timbre-similar and background-clean sound audio based on the location of events in the silent video and the acoustic characteristics in the reference audio.
The predicted mel-spectrogram will be passed through a pre-trained HifiGAN vocoder \cite{kong2020hifi} to get the final waveform.

To validate the effectiveness of our method, we conduct a series of experiments on the VAS dataset.
Since there are no evaluation metrics for this new task yet, we introduce three Mean Opinion Scores (MOS) as the subjective evaluation metrics which evaluate the generated audio in: the realism, the temporal alignment with video, and the timbre similarity with reference audio, respectively.
The evaluation of the generated results shows that our model can generate a high-quality audio with high temporal alignment with the input silent video and high timbre similarity with the reference audio though obtaining a score close to the recorded audio.

The main contributions of this paper are summarized as follows:
\begin{itemize}
    \item We define a new task called timbre-controlled video-to-audio generation for solving the problem of generating realistic sound effects with the desired timbre for silent videos.
    \item We propose a method based on information disentangle and adversarial training, with an efficient encoder-decoder structure for solving this problem.
    \item The results of our experiment achieve MOS scores close to the ground truth on the subjective evaluation, proving that we accomplished the proposed task effectively and with high quality.
\end{itemize}

%% file: Sections/2_related_works.tex
\section{Related Works}
\subsection{Video-to-Audio Generation}
A series of audio-visual related generation tasks have been proposed and produced some achievements.
\cite{owens2016visually} generates audio according to the video of hitting and scratching objects with a drumstick, which is the first to propose the task of video-to-audio generation as far as we know.
\cite{aytar2016soundnet,owens2016ambient,takahashi2017aenet}'s models and experiments successfully demonstrate modality and informational correlation between the video and audio.
\cite{zhou2018visual} extends video-to-audio generation tasks to the sound in the wild.
\cite{chen2018visually} uses an autoregressive approach to learn the corresponding sound from video features.
\cite{chen2020generating} and \cite{ghose2021foleygan} introduce generative adversarial networks based on previous work to obtain better audio quality.
\cite{SpecVQGAN_Iashin_2021} uses codebooks to modeling a better representation between the video and audio.

However, the above works only predict the corresponding audio from the video and do not get the timbre they want on demand.
Some of them still have significant problems with the quality of the generated audio, such as unclear sound or misalignment in time.

\subsection{Information Disentanglement}
Information Disentanglement aims to make all the latent distribution dimensions independent from each other, which is an approach to solving a diverse set of tasks by disentangling (or isolating) the underlying structure of the main problem (or object) into disjoint parts of its representations. 

Methods based on variational autoencoders are the most common practice in information disentanglement.
\cite{mathieu2019disentangling} develops a generalization of disentanglement in variational autoencoders by holding different axes of variation fixed during training.
\cite{hsieh2018learning} proposes a model that explicitly decomposes and disentangles the video representation and reduces the complexity of future frame prediction. 
\cite{carbajal2021disentanglement} uses an adversarial training scheme for variational autoencoders to disentangle the label from the other latent variables and achieve some progress in speech enhancement.
\cite{qian2020unsupervised} and \cite{qian2019autovc} employ bottlenecks to disentangle the information in speech audio and have achieved significant results in voice conversion tasks.


\subsection{Generative Adversarial Network}
Generative Adversarial Network \cite{goodfellow2014generative} is first proposed in the image generation tasks, which have achieved considerable success in generating high-quality images \cite{donahue2019large,karras2019style,zhang2022towards,huang2022singgan,huang2021multi,zhang2022adversarial}.

In the audio-related tasks, GAN has also achieved some progress in modeling realistic audio with the growth of audio generation recently. 
\cite{donahue2018adversarial} is the first work that introduces the GAN model into the unsupervised synthesis of raw-waveform audio.
\cite{sheng2019high} utilizes a generative enhancer model for mel-spectrogram that reduces the gap between the true mel-spectrogram and the generated mel-spectrogram to overcome the over-smooth in the generated results.
\cite{lee2021multi} is trained with frame-level adversarial feedback to synthesize high-fidelity mel-spectrograms by making the discriminator learn to distinguish which features are converted to mel-spectrogram with a frame-level condition.
\cite{yang2021ganspeech} applies adversarial training to the non-autoregressive text-to-speech model and achieves a good performance in multi-speaker speech-generating.

Besides directly performing adversarial training on the mel-spectrograms, some works \cite{kumar2019melgan,yamamoto2020parallel,kong2020hifi,huang2022transpeech,huang2022generspeech} use GAN to improve the capability of the model for generating from the mel-spectrograms to the waveform (i.e., vocoder).



%% file: Sections/3_method.tex
\section{VarietySound}

This section introduces the proposed method and the model architecture of the proposed VarietySound in detail.
\subsection{Overview}
\begin{figure}[t]
    \centering
    \begin{minipage}[c]{\linewidth}
    \centering
    \includegraphics[width=\linewidth]{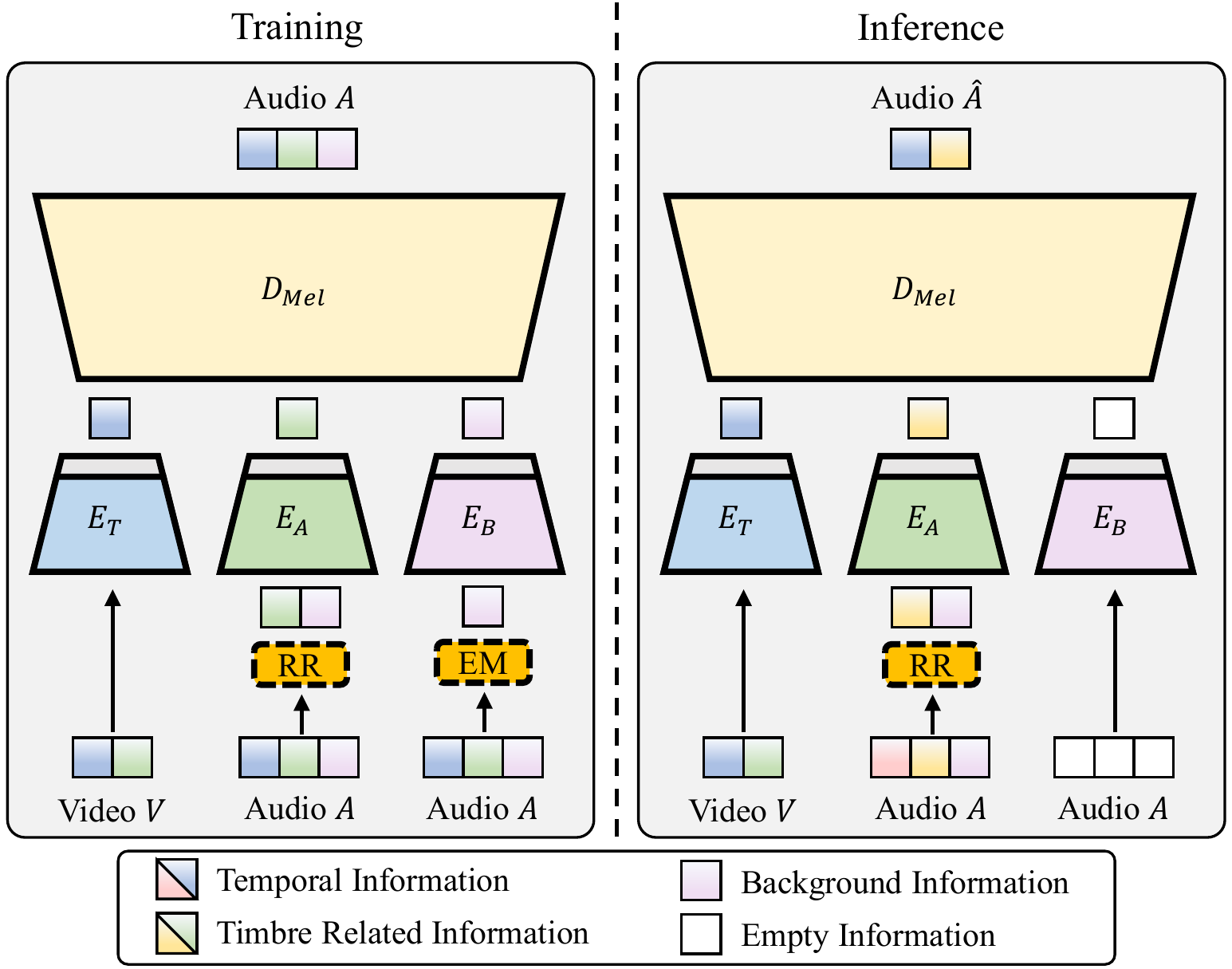}
    \end{minipage}%
    \vspace{-2mm}
    \caption{Information Disentanglement. $\boldsymbol{E_{T}}$, $\boldsymbol{E_{A}}$ and $\boldsymbol{E_{B}}$ denotes the Temporal Encoder, the Acoustics Encoder and the Background Encoder, respectively. $\boldsymbol{D_{mel}}$ denotes the Mel Decoder. $\boldsymbol{RR}$ denotes the random resampling transform \textbf{IN} the Acoustics Encoder and $\boldsymbol{EM}$ denotes Energy Masking Operation before the Background Encoder.}
    
    \label{fig:method}
    \vspace{-6mm}
\end{figure}

As shown in Figure \ref{fig:method}, our method is a process of information disentanglement and re-fusion.
We first disentangle the final audio information into three components: temporal information, timbre information, and background information (see Sec. \textit{Information Components}), modeling them with three different encoders respectively, and then use a mel decoder to recombine these disentangled information for the reconstruction of the audio.
The disruption operations and the bottlenecks force the encoders to pass only the information that other encoders cannot supply, hence achieving the disentanglement; and with the sufficient information inputs, the mel decoder could finish the reconstruction of the target mel-spectrogram, hence achieving the re-fusion.
We also adopt the adversarial training, which helps the model to fit the distribution of the target mel-spectrogram better and thus obtain higher quality and better temporal alignment generation results.

In the training phase of the generator, we use video features and mel-spectrogram from the same sample feed into the network, where the video features are fed into the Temporal Encoder and the mel-spectrogram is fed into the Acoustic and Background Encoders.
Our disentanglement method is unsupervised because there is no explicit intermediate information representation as a training target.
The outputs of the three encoders are jointly fed to the Mel Decoder to obtain the final reconstructed mel-spectrogram, and the generation losses are calculated with the real mel-spectrogram as Sec. \textit{Generator Loss} to guide the training of the generator.

In the training phase of the discriminator, for the Time-Domain Alignment Discriminator, the video features and mel-spectrogram from the same real sample are used as inputs to construct positive samples, while the real video features and the reconstructed mel-spectrogram are used as inputs to construct negative samples.
For the Multi-Window Mel Discriminator, the real mel-spectrogram from the sample is used as a positive sample and the reconstructed mel-spectrogram is used as a negative sample input.
The two discriminators calculate the losses and iteratively train according to the method in Sec. \textit{Discriminator Loss}.

In the inference phase, we feed the video features into the temporal encoder, the mel-spectrogram of the reference audio containing the target timbre into the acoustic encoder, and the mel-spectrogram of the muted audio into the background encoder, and then generate the sound through the mel decoder.
The choice of reference audio is arbitrary, depending on the desired target timbre.
Theoretically, the length of the video features and reference audio input during the inference phase is arbitrary, but it is necessary to ensure that the relevant events are present in the video and that the reference audio contains the desired timbre to obtain the normally generated sound.



\subsection{Information Components}
\label{sec:info}
\subsubsection{Temporal Information}
The temporal information refers to the location information in the time sequence corresponding to the occurrence of the sound event.
In the temporal sequence, the position where the sound event occurs strongly correlates with the visual information adjacent to that position for real recorded video.
Therefore, in our method, this part of the information will be predicted using the visual feature sequence of the input video.
We also set a suitable bottleneck to ensure that the video can provide only temporal information without providing other acoustic content information.

\subsubsection{Timbre Information}
Timbre information is considered an acoustic characteristic inherent to the sound-producing object.
The distribution of timbres between different categories of objects can vary widely, and the timbres of different individuals of the same category of objects usually possess specific differences.
In our method, this part of the information will be predicted by the reference audio.
The random resampling transform refers to the operations of segmenting, expand-shrink transforming and random swapping of the tensor in the time sequence.
When encoding the reference timbre information, we perform a random resampling transform on the input reference audio in the time sequence to disrupt its temporal information.

\subsubsection{Background Information}
Background information is perceived as timbre-independent other acoustic information, such as background noise or off-screen background sound.
This part of the information is necessary for training to avoid model confusion due to the information mismatch.
We found that the energy of this part of the information is usually much smaller in the mel-spectrogram than the part where the timbre information is present.
Therefore, in the proposed method, we adopt an Energy Masking operation that masks the mel-spectrogram of the part of the energy larger than the median energy of the whole mel-spectrogram along the time dimension.
The energy masking operation discards both temporal and timbre-related information of the mel-spectrogram, preserving only the background information in the audio.
In the training phase, this information is added to match the target mel-spectrogram; in the inference phase, this information will be set to empty to generate clearer audio.

\subsection{Training Losses}
In the proposed method, we use a generative adversarial structure for training, including one generator $G(\cdot)$ with a time-domain alignment discriminator $D_{t}(\cdot)$ and a mel discriminator $D_{m}(\cdot)$.
We denote the input video, reference audio and the generator's generate result of our model as $\boldsymbol{v}\sim p_{\text{data}}(\boldsymbol{v})$,$\boldsymbol{m}\sim q_{\text{data}}(\boldsymbol{m})$ and $G(\boldsymbol{v}, \boldsymbol{m})=\boldsymbol{\hat{m}}\sim q_{\text{generated}}(\boldsymbol{\hat{m}})$, respectively.

\subsubsection{Generator Loss}
\label{sec:gloss}
The generator loss $L_{\mathrm{G}}$ for training is defined by the following equation:
\begin{equation}
L_{\mathrm{G}}(G, D_{m}, D_{t}) = \lambda_{m} L_{\mathrm{mel}}+ \lambda_{a} L_{\mathrm{adv}}
\end{equation}
where $\lambda_{m}$ and $\lambda_{a}$ are hyper parameters for training and the definition of the mel-spectrogram loss $L_{mel}$ and the adversarial loss $L_{adv}$ is:
\begin{equation}
L_{\mathrm{mel}}= \|\boldsymbol{\hat{m}}-\boldsymbol{m}\|_{1}
\end{equation}

\begin{equation}
L_{\mathrm{adv}}=E_{v, \boldsymbol{\hat{m}}}\left[\left\|1-D_t(\boldsymbol{\hat{m}} \mid \boldsymbol{v})\right\|_2+\left\|1-D_m(\boldsymbol{\hat{m}})\right\|_2\right]
\end{equation}

\subsubsection{Discriminator Loss}
\label{sec:dloss}
The loss to minimize when training the time-domain alignment discriminator is defined by the following equation:
\begin{equation}
L_{\mathrm{D_{t}}}(G, D_{t}) = L_{\mathrm{real}}+ L_{\mathrm{fake}}
\end{equation}
and the definition of the positive sample loss $L_{real}$ and the negative sample loss $L_{fake}$ is: 

\begin{equation}
L_{\mathrm{real}}=E_{\boldsymbol{v}, \boldsymbol{m}}\left[\left\|1-D_t(\boldsymbol{m} \mid \boldsymbol{v})\right\|_2\right]
\end{equation}


\begin{equation}
L_{\mathrm{fake}}=E_{\boldsymbol{v},\boldsymbol{\hat{m}},\boldsymbol{m}}\left[\left\|-D_t(\boldsymbol{\hat{m}} \mid \boldsymbol{v})\right\|_2+\left\|-D_t(\mathbf{S}[\boldsymbol{\hat{m}}] \mid \boldsymbol{v})\right\|_2\right]
\end{equation}

where the $\mathbf{S}[\cdot]$ donates the shift operation to build the negative samples with time misaligned.

It is worth noting that we use two negative samples in training the time-domain alignment discriminator: a generated samples of the generator and a real samples after time-shifting.
The advantage of this approach is that we use these negative samples to increase the discriminator's ability to discriminate generated spectra that are consistent in content but not aligned in time, thus allowing the generator to model the temporal location of voicing more accurately.

The loss of the mel discriminator is defined by the following equation:

\begin{equation}
L_{\mathrm{D}_{\mathrm{m}}}\left(G, D_m\right)=E_{\boldsymbol{\hat{m}}}\left[\left\|1-D_m(\boldsymbol{m})\right\|_2+\left\|-D_m(\boldsymbol{\hat{m}})\right\|_2\right]
\end{equation}

\subsection{Model Architecture}
\begin{figure*}[htbp]
    \captionsetup[subfigure]{justification=centering}
    \centering
    \subfigure[Overview]{
        \begin{minipage}[b]{0.26\linewidth}
        \centering
        \includegraphics[width=\linewidth]{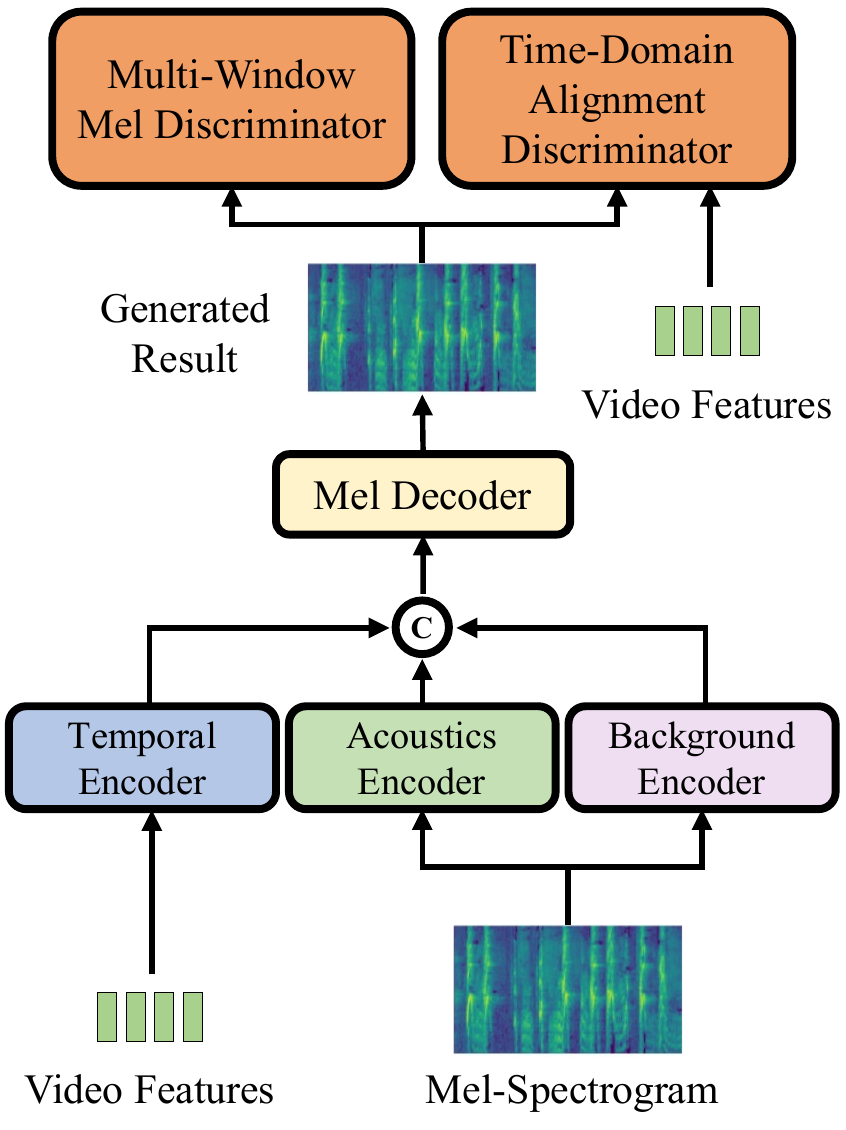}
        \label{fig:model}
        \vspace{-4mm}
        \end{minipage}%
    }
    \subfigure[TE]{
        \begin{minipage}[b]{0.102\linewidth}
        \centering
        \includegraphics[width=\linewidth]{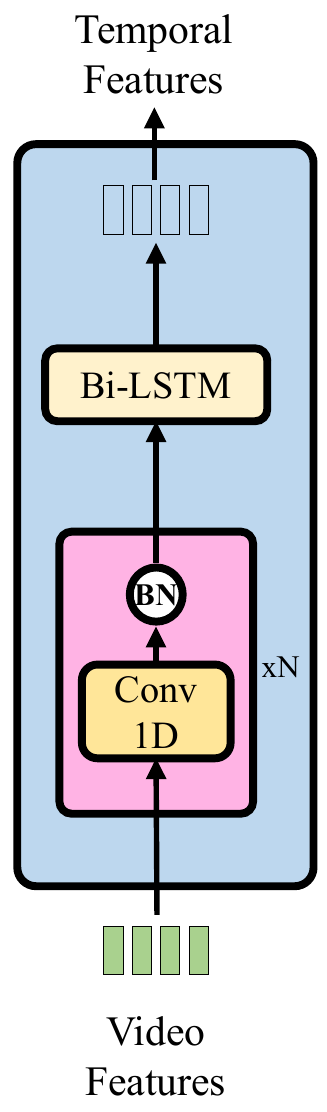}
        \label{fig:tdr}
        \vspace{-4mm}
        \end{minipage}%
    }%
    \subfigure[AE] {
        \begin{minipage}[b]{0.125\linewidth}
        \centering
        \includegraphics[width=\linewidth]{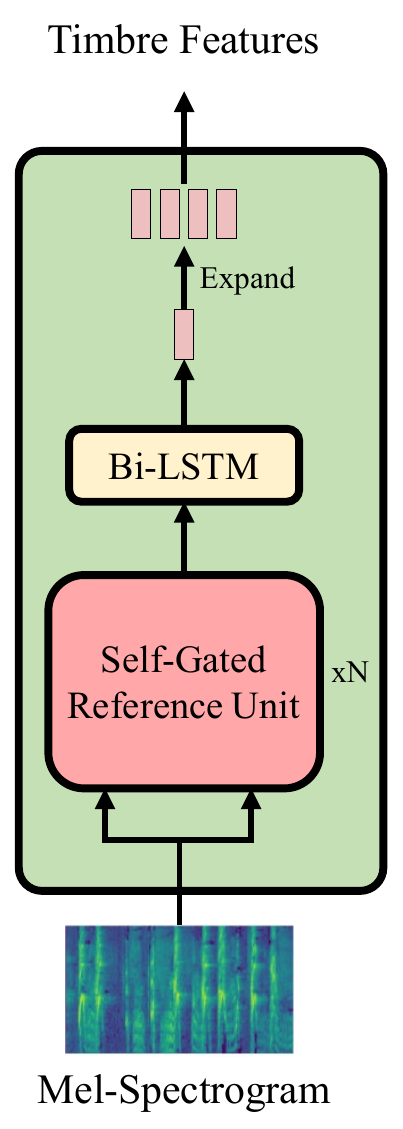}
        \label{fig:ref}
        \vspace{-4mm}
        \end{minipage}%
    }%
    \subfigure[BE]{
        
        \begin{minipage}[t]{0.12\linewidth}
        \centering
        \includegraphics[width=0.845\linewidth]{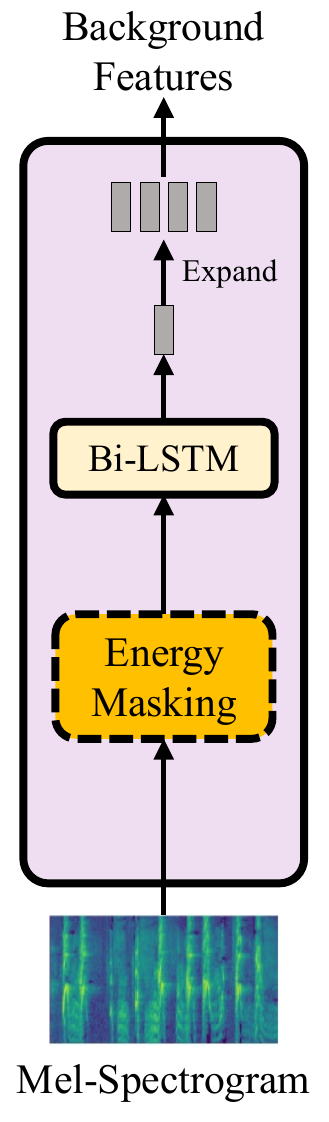}
        \label{fig:bg}
        \vspace{-4mm}
        \end{minipage}%
    }%
    \subfigure[Mel Decoder]{
        \begin{minipage}[b]{0.17\linewidth}
        \centering
        \includegraphics[width=\linewidth]{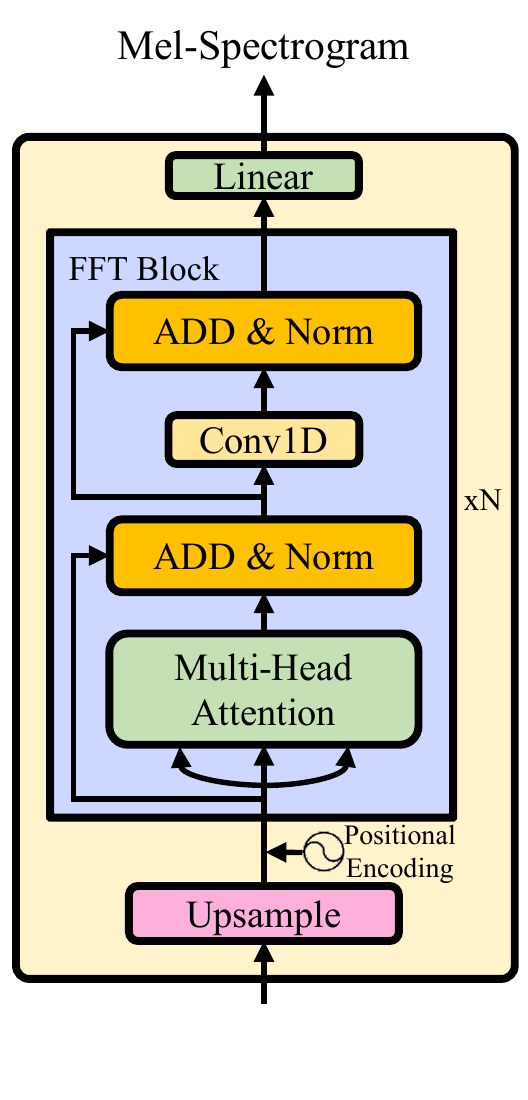}
        \label{fig:dec}
        \vspace{-4mm}
        \end{minipage}%
    }%
    \subfigure[TDAD]{
        \begin{minipage}[b]{0.155\linewidth}
        \centering
        \includegraphics[width=\linewidth]{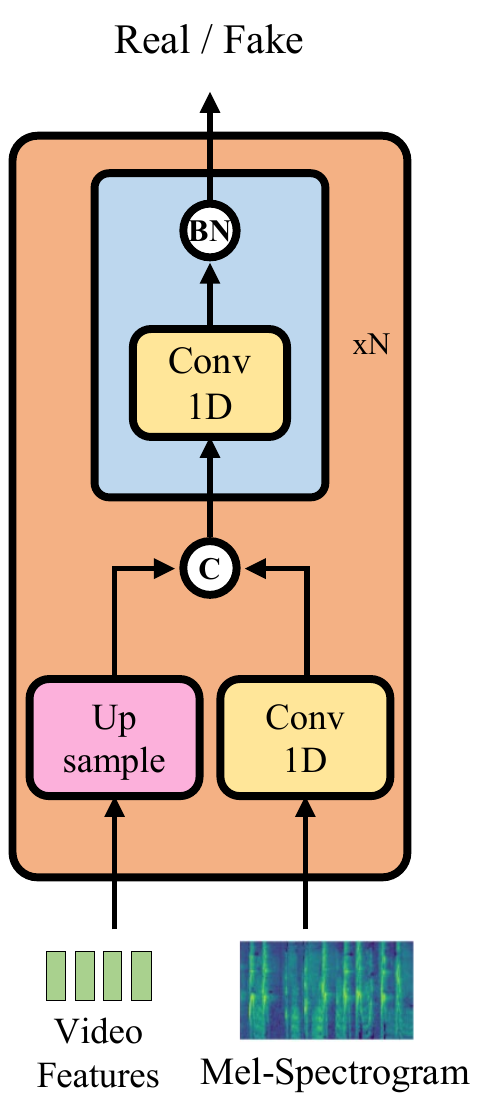}
        \label{fig:disc}
        \vspace{-4mm}
        \end{minipage}%
    }%
    \vspace{-2mm}
    \caption{Model architecture. ``TE" denotes the Temporal Encoder, ``AE" denotes the Acoustic Encoder, ``BE" denotes the Background Encoder and ``TDAD" denotes the Time-Domain Alignment Discriminator. \copyright\space denotes the concatenation.}
    \vspace{-2mm}
\end{figure*}
The model structure is shown in the Figure \ref{fig:model}. The generator consists of four parts, including three encoders and one decoder.
The three encoders modeling the three parts of information in our proposed method separately and the decoder uses the information encoded by the encoders for the reconstruction of the mel-spectrogram.
In addition, to discriminate the temporal alignment and the quality of the generated mel-spectrogram, we introduce two discriminators: a Time-Domain Alignment Discriminator and a Multi-Window Mel discriminator.

\subsubsection{Temporal Encoder}
The temporal encoder modeling the temporal location of the acoustic event from the video feature sequence.
Specifically, the temporal encoder accepts video sequence features as input and encodes hidden sequences with the same length as the video features.
As shown in Figure \ref{fig:tdr}, the temporal encoder consists of a multi-layer 1D convolutional stack and uses bidirectional LSTM layers to encode the temporal sequence information after the convolution layers.
Each 1D convolutional layer is activated using leaky ReLU and normalized with the Batch Normalization.
Finally, a linear layer is employed as a bottleneck layer to limit the information.
The output of the temporal encoder is known as the Temporal Features.

\subsubsection{Acoustic Encoder}
We propose an acoustic encoder for encoding the timbre information which is required to reconstruct the mel-spectrogram.
This encoder consists of a multi-layer Self-Gated acoustic unit (SGAU) and a single layer of the bidirectional LSTM as shown in Figure \ref{fig:ref}.
Inspired by the GAU in WaveNet \cite{oord2016wavenet}, we designed this gated structure to extract the timbre information from the mel-spectrogram.
Each SGAU accepts two inputs and gives two outputs what we call feature input/output and condition input/output, the detailed description and the structure of the SGAU are added to \textit{Appendix A}\ref{ap:SGAU}.
In our architecture, the first layer of the SGAUs has the same vector at both inputs and that is why we call it ``Self-Gated".
After the SGAUs, a bidirectional LSTM layer is added for dimensionality reduction along the temporal dimension.
We use the hidden vector of the last layer of the bidirectional LSTM as the timbre features and expand it to the same length as the temporal features in the temporal dimension.

\subsubsection{Background Encoder}
The background encoder is used to encode the acoustic independent background information required to reconstruct the mel-spectrogram.
As shown in Figure \ref{fig:bg}, the background encoder consists of an energy masking operation, a bidirectional LSTM layer and a linear layer.
Similar to the operation in the acoustic encoder, we use the hidden vector of the last layer of the bidirectional LSTM as the background features, which is expanded to the same length as the temporal features after dimensionality reduction from the linear layer.
The input of the background encoder will be set to zero (i.e., silent audio) in the inference stage.

\subsubsection{Mel Decoder}
As shown in Figure \ref{fig:dec}, the outputs of the three encoders are concatenated along the time dimension and then fed to a mel decoder to generate the final mel-spectrogram in parallel.
In order to upsample to the same length as the mel-spectrogram in the time dimension, the input hidden vector is first fed to an upsampling layer which consists of a series of 1D transposed convolutional layers.
After that, the upsampled sequences are position coded and then passed through several layers of feed-forward Transformer (FFT) Block, which is used as Mel decoder in FastSpeech 2 \cite{ren2020fastspeech}.
Each FFT block consists of a self-attention and 1D convolutional network.
Following Transformer \cite{vaswani2017attention}, residual connections, layer normalization, and dropout are added after the self-attention network and 1D convolutional network, respectively.
Finally, the feature dimension of the sequence is transformed to the number of mel bins by a single linear layer to get the mel-spectrogram.

\subsubsection{Time-Domain Alignment Discriminator}
The Time-Domain Alignment Discriminator (TDAD) is a conditional discriminator.
As shown in the Figure \ref{fig:disc}, the discriminator receives two inputs: the mel-spectrogram as the discriminate target and the video feature sequence as the condition.
In order to keep consistent with the mel-spectrogram in the time dimension, the video feature sequences are upsampled through an upsampling layer. 
The mel-spectrogram features are concatenated with the upsampled video features after a 1D convolution layer and fed into a multi-layer 1D convolutional structure for dimensionality reduction.
The Batch Normalization and leaky ReLU activation will be added after the 1D convolutional layer except for the last layer.
Before the output, we apply sigmoid activation to the result.

\subsubsection{Multi-Window Mel Discriminator}
We also use a Multi-Window Mel Discriminator to discriminate the quality of the generated mel-spectrogram.
The Multi-Window Mel discriminator shares the same structure in SyntaSpeech \cite{ye2022syntaspeech} which consists of 3 sub-discriminators with three 2D convolutional layers in each sub-dicriminator.


%% file: Sections/5_experiments.tex
\section{Experiments and Evaluation}
\begin{table*}[htbp]
    \centering

    \vspace{-2mm}   
    \begin{tabular}{c|ccc|ccc}
        \toprule
        \multirow{2}{*}{\textbf{Category}} &
        \multicolumn{3}{c|}{\textbf{Audio Realism}} &
        \multicolumn{3}{c}{\textbf{Temporal Alignment}}\\
        & \multicolumn{1}{c|}{GT}& \multicolumn{1}{c|}{Baseline} & \multicolumn{1}{c|}{Ours} & \multicolumn{1}{c|}{GT}& \multicolumn{1}{c|}{Baseline} & \multicolumn{1}{c}{Ours}   \\
        \midrule
        Baby & 4.55 $(\pm 0.10)$ &2.67 $(\pm 0.23)$& 3.77 $(\pm 0.15)$ & 4.43 $(\pm 0.08)$\ &3.83 $(\pm 0.17)$& 4.07 $(\pm 0.10)$ \\
        Cough & 4.32 $(\pm 0.11)$ &3.30 $(\pm 0.20)$& 4.13 $(\pm 0.12)$ & 4.30 $(\pm 0.11)$ &3.71 $(\pm 0.24)$& 4.17 $(\pm 0.12)$ \\
        Dog & 4.45 $(\pm 0.11)$ &3.21 $(\pm 0.19)$& 4.18 $(\pm 0.11)$ & 4.45 $(\pm 0.08)$ &4.32 $(\pm 0.15)$& 4.40 $(\pm 0.08)$ \\
        Drum & 4.62 $(\pm 0.08)$ &2.91 $(\pm 0.21)$& 4.12 $(\pm 0.15)$ & 4.56 $(\pm 0.06)$ &3.64 $(\pm 0.16)$& 4.25 $(\pm 0.11)$ \\
        Fireworks & 4.56 $(\pm 0.09)$ &3.16 $(\pm 0.22)$& 4.23 $(\pm 0.13)$ & 4.47 $(\pm 0.08)$ &4.00 $(\pm 0.21)$& 4.35 $(\pm 0.10)$  \\
        Gun & 4.38 $(\pm 0.12)$ &2.76 $(\pm 0.22)$& 4.02 $(\pm 0.15)$ & 4.45 $(\pm 0.09)$ &4.08 $(\pm 0.17)$& 4.25 $(\pm 0.12)$ \\
        Hammer & 4.43 $(\pm 0.12)$ &3.16 $(\pm 0.26)$& 3.84 $(\pm 0.14)$ & 4.31 $(\pm 0.08)$ &3.88 $(\pm 0.19)$& 4.19 $(\pm 0.13)$ \\
        Sneeze & 4.04 $(\pm 0.13)$ &2.75 $(\pm 0.22)$& 4.00 $(\pm 0.15)$ & 4.28 $(\pm 0.12)$ &3.76 $(\pm 0.23)$& 4.16 $(\pm 0.11)$ \\ 
        \midrule
        \textbf{Average} & $\mathbf{4.42(\pm 0.04)}$ &2.99 $(\pm 0.08)$& $\mathbf{4.04(\pm 0.05)}$ & $\mathbf{4.41(\pm 0.03)}$ &3.90 $(\pm 0.07)$& $\mathbf{4.23(\pm 0.04)}$ \\
        \bottomrule
        
    \end{tabular}
    \caption{Subjective Evaluation Results of Audio Realism and Temporal Alignment. }
    \label{tab:Evaluation}
\end{table*}
In this section, we will describe in detail how our experiments are performed and give our evaluation results.

\subsection{Experimental Setup}
\subsubsection{Dataset and Processing}
We evaluate our method and model on the VAS dataset \cite{chen2020generating}, which is proposed for audio-visual related machine learning tasks.
VAS has a total of 12541 videos in eight categories (\textit{Baby, Cough, Dog, Drum, Fireworks, Gun, Hammer} and \textit{Sneeze}), whose sound is highly synchronized with the visual content.
We split the dataset into three sets for each video category: 32 samples for validation, 32 samples for testing and all the remaining samples for training.
We refer to the method in REGNET \cite{chen2020generating} for the initial processing of the data, using the \textit{ffmpeg} tool\footnote{\url{https://www.ffmpeg.org/}} to separate the audio and video of all raw data.
To facilitate video feature extraction, we convert the video data to 256 pixels in height by 340 pixels in width and limit the duration to 10 seconds by padding and clipping, while the frame rate is set to 21.5 frames per second (fps).
For the audio data, we unify the sampling rate to 22050Hz and unify the duration to 10 seconds.
After that, we perform feature extraction for video and audio data separately.
Specifically, a BN-Inception \cite{ioffe2015batch} model is utilized as a feature extractor to explore RGB and optical flow features at T different time steps.
The concatenation of RGB and optical flow feature sequences will be used as the input video features during the training stage.
As for the audio data, we apply a Short-Time Fourier Transform (STFT) on the original waveform and convert it to an 80-dimensional mel scale through a Mel Filter Bank.
The frame size and hop size of the STFT are set to 1024 and 256, respectively.
As a result, each video is processed into a vector with the shape of 2048*215, and its corresponding audio is processed into a mel-spectrogram with the shape of 80*860.

\subsubsection{Training and Inference}
We train our VarietySound on single NVIDIA GeForce RTX 3090 GPU, with batchsize of 48 samples.
We use the AdamW optimizer \cite{loshchilov2018decoupled} with $\beta_{1}=0.5$, $\beta_{2}=0.999$ and $\epsilon=10^{-8}$ and follow the same learning rate schedule in Regnet.
We train 500 epochs in each category respectively to obtain the final results.
During the inference, the output 80-dimensional mel-spectrograms of our model are transformed into waveform audio by a pre-trained HifiGAN Vocoder \cite{kong2020hifi}.

\subsubsection{Model Configuration}
Our VarietySound consists of 5 SGAUs in the acoustic encoder.
The number of the 1D convolutional layers in the Temporal Encoder and the Time-Domain Alignment Discriminator is set to 8 and 4, respectively.
In the mel-spectrogram decoder, there are 4 FFT blocks in total to be used.
We add more detailed configurations of our VarietySound used in our experiments in \textit{Appendix B}\ref{ap:conf}.

\subsubsection{Baseline Model}
Since we are the first to propose this task (i.e., TCVSG), there is no suitable baseline model that can provide comparable results for our experiments, and therefore to demonstrate the effectiveness of our model, we use a cascade model as the baseline to compare with. 
We add more details about the baseline model in \textit{Appendix C}.

\subsubsection{Evaluation Metrics}


Since there are no exists evaluation metrics for proposed tasks, we first defined a series of human tests for the subjective evaluation.
We use Mean Opinion Score (MOS) as the subjective evaluation metrics in three areas: the quality of the generated audio, the temporal alignment with the video, and the similarity in timbre to the reference audio. We describe in detail the definitions for the three scores and how our sample was selected in the \textit{Appendix D}\ref{ap:mos}.
In addition, in order to show the outstanding ability of the proposed model in modeling timbre information more objectively, we used the cosine similarity of the acoustic features as an objective evaluation of the similarity in timbre.
In order to exclude interference from the quality of the vocoder, the ground truth audios are generated by the vocoder using the ground truth mel-spectrogram in all the subject evaluations.

\begin{table*}[htbp]
    \centering

    \vspace{-2mm}   
    \begin{tabular}{c|cc|cc}
        \toprule
        \multicolumn{5}{c}{\textbf{Timbre Similarity}}\\
        \hline
        \multirow{2}{*}{\textbf{Category}} &

         \multicolumn{2}{c|}{\textbf{MOS}}&
        \multicolumn{2}{c}{\textbf{Cosine Similarity}}
        \\
        & \multicolumn{1}{|c|}{Baseline} & \multicolumn{1}{c|}{Ours} & \multicolumn{1}{c|}{Baseline} & \multicolumn{1}{c}{Ours}  \\
        \midrule
        Baby & 3.46 $(\pm 0.17)$& 3.94 $(\pm 0.08)$& 0.86 $(\pm 0.01)$& 0.88 $(\pm 0.00)$\\
        Cough &3.48 $(\pm 0.22)$& 3.59 $(\pm 0.09)$& 0.86 $(\pm 0.00)$& 0.93 $(\pm 0.01)$\\
        Dog & 3.63 $(\pm 0.15)$& 4.09 $(\pm 0.08)$& 0.77 $(\pm 0.00)$& 0.96 $(\pm 0.00)$\\
        Drum & 3.72 $(\pm 0.13)$& 3.85 $(\pm 0.09)$& 0.74 $(\pm 0.03)$& 0.84 $(\pm 0.00)$\\
        Fireworks &3.43 $(\pm 0.25)$& 3.93 $(\pm 0.07)$& 0.88 $(\pm 0.01)$& 0.89 $(\pm 0.01)$\\
        Gun & 3.45 $(\pm 0.18)$& 3.98 $(\pm 0.08)$& 0.83 $(\pm 0.01)$& 0.88 $(\pm 0.01)$\\
        Hammer & 3.74 $(\pm 0.17)$& 3.99 $(\pm 0.10)$& 0.80 $(\pm 0.02)$& 0.89 $(\pm 0.02)$\\
        Sneeze & 3.62 $(\pm 0.18)$& 3.72 $(\pm 0.08)$& 0.88 $(\pm 0.01)$& 0.98 $(\pm 0.01)$\\ 
        \midrule
        \textbf{Average} & 3.57 $(\pm 0.07)$& $\mathbf{3.89 (\pm 0.03)}$ & 0.84 $(\pm 0.01)$& $\mathbf{0.90 (\pm 0.01)}$\\
        \bottomrule
        
    \end{tabular}
    \caption{Evaluation Results of Timbre Similarity. }
    \label{tab:sim}
\end{table*}
\subsection{Experiment Results}

\subsubsection{Evaluation Results}
Through the third-party evaluation on the \textit{Amazon Mechanical Turk} (AMT), we obtained the evaluation results of our model. We also give several cases of mel-spectrogram comparison of our generated result and the ground truth in \textit{Appendix F}\ref{ap:res}, and you can also watch the demo videos on our demo page\footnote{\url{https://conferencedemos.github.io/Anonymous9517/}}.

As shown in the Table \ref{tab:Evaluation}, the proposed model achieves scores closer to ground truth in terms of both audio realism and temporal alignment by comparing with the baseline model.
The category of \textit{Dog} and \textit{Fireworks} have the best average performance in the two evaluations.
The category of \textit{Baby} gains the worst performance in the evaluation of audio realism and temporal alignment due to the uncertainty and diversity in human behavior which is hard for modeling, the same trend also appears in the category of \textit{Cough} and \textit{Sneeze}.
Due to the imbalance in the amount of data in each category in the dataset, we can see that the four categories with smaller amounts of data (\textit{Cough, Gun, Hammer} and \textit{Sneeze}) will have overall lower temporal alignment scores than the four categories with larger amounts of data (\textit{Baby, Dog, Fireworks} and \textit{Drum}) in both evaluations, suggesting that the modeling of temporal alignment may be more sensitive to the amount of data.

In the evaluation of the audio quality, the baseline model achieved a relatively low score. 
This is because the cascade model accumulates the errors of both models during the generation process, bringing apparent defects to the generated audio, such as noise, electrotonality, or mechanicalness.

For the similarity of the timbre, as shown in Table \ref{tab:sim}, the proposed model achieve higher scores both in the subjective and objective evaluation, which means the result of proposed model have a timbre closer to the ground truth than the baseline model.

We did not compare the generation speed because, empirically, the inference efficiency of a single model is usually much higher than that of a cascade model.

As a summary, by obtaining generation results and subjective evaluation results above, we have successfully demonstrated the effectiveness of our method and model.

\subsubsection{Ablation Study}

To demonstrate the effectiveness of all components in our model, we conducted several ablation experiments. We evaluated the generation results with the generator encoder(s) removed or with the discriminator removed, respectively, using the MCD \cite{kubichek1993mel} scores as a criterion. The MCD scores show the overall difference by calculating the Euclidean distance between the mel-spectrogram. In this criterion, the generated result is closer to the ground truth mel-spectrogram when it has a lower score. As a result, the generated samples of our model acquires the minimum MCD score on average, which has successfully demonstrate the usefulness of the three encoders for information disentanglement and the two discriminators for the adversarial training. We have included more specific results and analysis in the \textit{Appendix E}\ref{ap:abl}.

%% file: Sections/6_conclusion.tex
\section{Conclusion}

In this paper, for the problem of generating audio with a specific timbre for silent video, we define a new task called Timbre Controllable Video to Sound Generation (TCVSG).
In addition, to accomplish this task, we propose a method based on information disentanglement, which first disentangles the target audio information into temporal, acoustic and other background information and then re-fusion them to reconstruct the mel-spectrogram.
Furthermore, according to our method, we propose a GAN-based model architecture called VarietySound, which can accept a video and a reference audio as inputs and generate high-quality, video time-aligned audio with the reference timbres.
Finally, we have successfully demonstrated the effectiveness of our method and model through a series of experiments and evaluations of several aspects.
The evaluation results show that the quality of the generated audio is close to the real audio while having a high timbre similarity with the input reference audio and a highly temporal alignment with the input silent video.
As future directions, we will seek to update our method based on the current model to accomplish cross-category timbre transfer and end-to-end generating during the video to audio generation.

%% file: Sections/7_appendix.tex
\newpage
\clearpage
\appendix

\section{A. Self-Gated Acoustic Unit}
\label{ap:SGAU}
\begin{figure}[H]
    \centering
    \vspace{-2mm}
    \begin{minipage}[c]{0.5\linewidth}
    \centering
    \includegraphics[width=\linewidth]{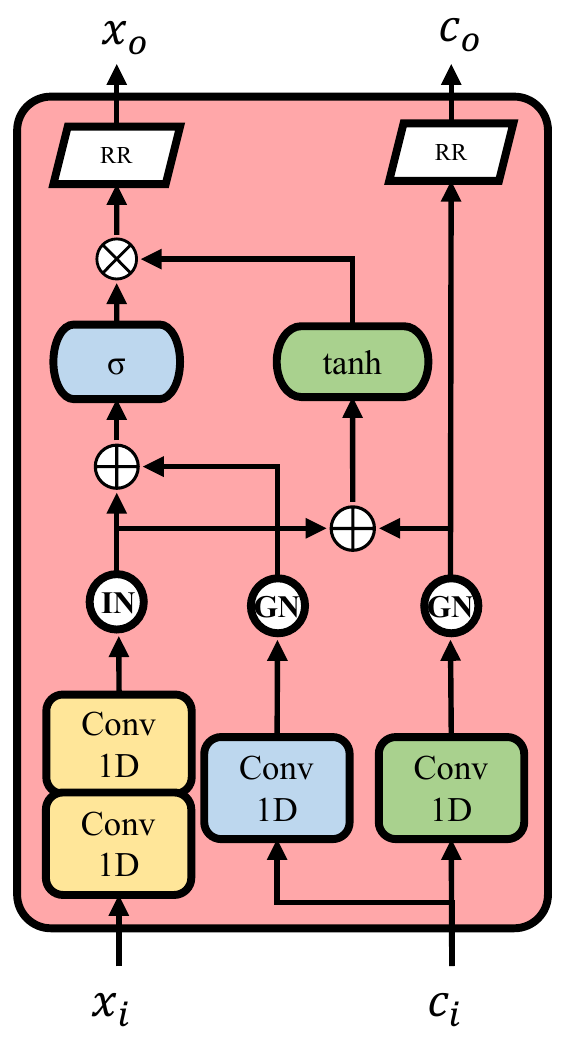}
    \end{minipage}%
    \vspace{-2mm}
    \caption{Self-Gated Acoustic Unit}
    \label{fig:SGAU}
    \vspace{-4mm}
\end{figure}
The structure of the SGAU is shown in Figure \ref{fig:SGAU}.
Each SGAU has two inputs and two outputs, which we call feature inputs, feature outputs, conditional inputs, and conditional outputs, respectively.
The feature input receives the input vectors and passes through two layers of 1D convolutional layers, which we call the input gate, and then normalized by Instance Normalization.
The conditional inputs receive the input vectors and then pass through two single 1D convolutional layers, which we call the output gate and skip gate.
The output gate and the skip gate are both normalized by Group Normalization.
The output of the jump gate is used as the output vector of the conditional output after the random resampling transform.
Meanwhile, the output of the input gate is added with the output of the skip gate and the output gate, respectively, and then multiplied after different activation functions.
The above result is transformed by Random Resampling and used as the output vector of the feature output.
For a clearer statement, the gated unit is described by the following equations:
\begin{equation}
    \mathbf{x_{o}}=\boldsymbol{R}[\tanh(\boldsymbol{W_{s}} * \mathbf{c_{i}}+\boldsymbol{V_{i}} * \mathbf{x_{i}}) \odot \sigma(\boldsymbol{W_{o}} *  \mathbf{c_{i}}+\boldsymbol{V_{i}} * \mathbf{x_{i}})]
\end{equation}
\begin{equation}
    \mathbf{c_{o}}=\boldsymbol{R}[\boldsymbol{W_{s}} * \mathbf{c_{i}}]
\end{equation}
where $\mathbf{x_{i}}$ and $\mathbf{c_{i}}$ denote two inputs of the unit, $\mathbf{x_{o}}$ and $\mathbf{c_{o}}$ denote two outputs of the unit. $\odot$ denotes an element-wise multiplication operator, $\sigma(\cdot)$ is a sigmoid function. $\boldsymbol{R}[ \cdot ]$ denotes the random resampling transform, $\boldsymbol{W_{ \cdot }}* $ and $\boldsymbol{V_{ \cdot }}* $ denote the single layer convolution in skip or output gate and the 2-layer convolutions in input gate separately.

\newpage
\section{B. Model Configuration}
\label{ap:conf}
We list hyperparameters and configurations of all models used in our experiments in Table \ref{tab:conf}.
\begin{table}[H]
    \vspace{-3mm}
    \centering

    \begin{tabular}{l c}
        \toprule
        \multicolumn{1}{c}{\textbf{Hyperparameter}} &  
        \multicolumn{1}{c}{\textbf{Size}}  \\
        \midrule
        Temporal Encoder Input Dimension & 2048\\
        \midrule
        Temporal Encoder Conv1D Layers & 8\\
        \midrule
        Temporal Encoder Conv1D Kernel & 5\\
        \midrule
        Temporal Encoder Conv1D Filter Size & 512\\
        \midrule
        Temporal Encoder LSTM Layers & 2\\
        \midrule
        Temporal Encoder LSTM Hidden & 256\\
        \midrule
        Temporal Encoder Output Dimension & 8 \\
        \midrule
        Acoustic Encoder SGAU Layers & 5\\
        \midrule
        SGAU Input-Gate Conv1D-1 Kernel & 5\\        
        \midrule
        SGAU Input-Gate Conv1D-2 Kernel & 7\\        
        \midrule
        SGAU Input-Gate Conv1D-(1\&2) Filter Size & 512\\        
        \midrule
        SGAU Output-Gate Conv1D Kernel & 3\\        
        \midrule
        SGAU Output-Gate Conv1D Filter Size & 512\\        
        \midrule
        SGAU Skip-Gate Conv1D Kernel & 5\\        
        \midrule
        SGAU Skip-Gate Conv1D Filter Size & 512\\         
        \midrule
        Acoustic Encoder LSTM Layers & 2\\        
        \midrule
        Acoustic Encoder LSTM Hidden & 256\\
        \midrule
        Background Encoder LSTM Layers & 2\\        
        \midrule
        Background Encoder LSTM Hidden & 128\\
        \midrule
        Mel Decoder ConvT1D Layers & 2\\
        \midrule
        Mel Decoder ConvT1D Kernel & 4\\        
        \midrule
        Mel Decoder ConvT1D Stride & 2\\        
        \midrule
        Mel Decoder ConvT1D Filter Size & 1024\\
        \midrule
        Mel Decoder FFT Blocks & 4\\
        \midrule
        FFT Block Hidden & 512\\
        \midrule
        FFT Block Attention Headers & 2\\
        \midrule
        FFT Block Conv1D Kernel & 9\\
        \midrule
        FFT Block Conv1D Filter Size & 512\\ 
        \midrule
        TDAD ConvT1D Layers & 2\\
        \midrule
        TDAD ConvT1D Kernel & 4\\        
        \midrule
        TDAD ConvT1D Stride & 2\\        
        \midrule
        TDAD ConvT1D Filter Size & 1024\\   
        \midrule
        TDAD Conv1D Layers & 4\\
        \midrule
        TDAD Conv1D Kernel & 4\\    
        \midrule
        TDAD Conv1D Filter Size & 512\\
        \midrule
        $\lambda_{m}$ & 10000.0\\
        \midrule 
        $\lambda_{a}$ & 1.0\\
        \bottomrule
    \end{tabular}
    \caption{Model Configuration}
    \vspace{-3mm}       
    \label{tab:conf}
    \vspace{-8mm}
\end{table}

\newpage

\section{C. Baseline Model}
\label{ap:base}
Specifically, we build our cascade model using a video-to-sound generation model and a sound conversion model. The video-to-sound generation model is responsible for generating the corresponding audio for the muted video, while the sound conversion model is responsible for converting the timbre of the generated audio to the target timbre.
We chose REGNET \footnote{\url{https://github.com/PeihaoChen/regnet/}} as the video-to-sound generation model, which has an excellent performance in the previous tasks.
For the sound conversion model, we consider using the voice conversion model which is used to accomplish similar tasks, since there is no explicitly defined sound conversion task and model.
We conducted some tests and found that some voice conversion models can perform simple sound conversion tasks.
Eventually, we chose unsupervised SPEECHSPLIT \footnote{\url{https://github.com/auspicious3000/SpeechSplit}} as the sound conversion model because of the lack of detailed annotation of the timbres in each category in the dataset.
The cascade model is trained on the same dataset (VAS) and in the same environment, and inference is performed using the same test data as in Sec.\textit{Sample Selection} \ref{sec:sample}.
In particular, instead of using speaker labels, our sound conversion model uses a sound embedding obtained from a learnable LSTM network as an alternative for providing target timbre information.
The cascade model's configuration follows the official implementation of the two models.

\section{D. Evaluation Design and Sample Selection}
\subsection{Evaluation Design}
We give the detailed definition of the MOS score on the subjective evaluation of audio realism, temporal alignment and timbre similarity in Table \ref{tab:real}, \ref{tab:tem} and \ref{tab:sim}, respectively.

In the evaluation of the realism of the generated audio, we will ask the raters to listen to several test audios and rate the realistic level of the audio content.
The higher the score, the more realistic the generated audio.

In the evaluation of temporal alignment, we ask the raters to watch several test videos with their audio and rate the alignment of them.
Samples with a shorter interval between the moment of the event in the video and the moment of the corresponding audio event will receive a higher score.

In the evaluation of timbre similarity, we ask the rater to listen to one original audio and several test audios and score how similar the test audio timbre is to the original audio timbre.
For cosine similarity, we calculate the cosine similarity between the target audio and ground truth audio timbre features using the following equation:
\begin{equation}
     CosSim(X,Y) = \frac{ \sum \limits_{i=1}^{n}(x_{i} * y_{i})}{ \sqrt{ \sum \limits_{i=1}^{n}(x_{i})^{2}} \sqrt{  \sum \limits_{i=1}^{n}(y_{i})^{2} } } 
\end{equation}
, and the timbre features are calculated by the third-party library\footnote{https://github.com/resemble-ai/Resemblyzer}.
The higher the similarity between the test audio and the original audio sound, the higher the score will be.
\begin{table}[ht]
    \centering
    \begin{tabular}{c c}
        \toprule
        \multicolumn{1}{c}{\textbf{Score}} &  
        \multicolumn{1}{c}{\textbf{Meaning}}  \\
        \midrule
         \textbf{5} &  Completely real sound.\\
        \midrule
        \textbf{4} & Mostly real sound.\\
        \midrule
        \textbf{3} & Equally real and unreal sound.\\
        \midrule
        \textbf{2} & Mostly unreal sound.\\
        \midrule
        \textbf{1} & Completely unreal sound.\\
        \bottomrule
    \end{tabular}
    \caption{MOS of Realism}
    \label{tab:real}
\end{table}
\begin{table}[ht]
    \centering
    \begin{tabular}{c c}
        \toprule
        \multicolumn{1}{c}{\textbf{Score}} &  
        \multicolumn{1}{c}{\textbf{Meaning}}  \\
        \midrule
         \textbf{5} & Events in video and events in audio\\& occur simultaneously.\\
        \midrule
        \textbf{4} & Slight misalignment between events\\& in video and events in audio.\\
        \midrule
        \textbf{3} & Exist misalignment in some positions.\\
        \midrule
        \textbf{2} & Exist misalignment in most of the positions.\\
        \midrule
        \textbf{1} & Completely misalignment, no events in audio\\& can match the video .\\
        \bottomrule
    \end{tabular}
    \caption{MOS of Temporal Alignment}
    \label{tab:tem}
\end{table}
\begin{table}[ht]
    \centering
    \begin{tabular}{c c}
        \toprule
        \multicolumn{1}{c}{\textbf{Score}} &  
        \multicolumn{1}{c}{\textbf{Meaning}}  \\
        \midrule
         \textbf{5} &  Timbre is exactly the same as target.\\
        \midrule
        \textbf{4} & Timbre has high similarity\\& with the target but not same.\\
        \midrule
        \textbf{3} & Timbre has similarity with the target,\\& but there are obvious differences.\\
        \midrule
        \textbf{2} & Timbre has a large gap with the target,\\& but share the same category of the sound.\\
        \midrule
        \textbf{1} & Timbre is completely different to the target.\\
        \bottomrule
    \end{tabular}
    \caption{MOS of Timbre Similarity}
    \label{tab:sim}
\end{table}

\subsection{Sample Selection}
\label{sec:sample}
To perform the evaluation, we randomly obtain test samples for each category in the following way.
Since our model accepts a video and a reference audio as a sample for input, we refer to it as a video-audio pair, and if the video and the audio are from the same raw video data, it will be called the original pair.

For audio realism evaluation, we will randomly select 10 data samples in the test set.
After breaking their original video-audio pairing relationship by random swapping, the new pair will be fed into the model to generate 10 samples, and then mixed these generated samples with the 10 ground truth samples to form the test samples.

For the temporal alignment evaluation, we will use the same method as above to obtain the generated samples and use the \textit{ffmpeg} tool to merge the audio samples with the corresponding video samples to produce the video samples with audio tracks.
We also mix 10 generated samples and 10 ground truth samples to form the test video samples in the temporal alignment evaluation.

For the timbre similarity evaluation, we randomly select 5 data samples in the test set, and for each audio sample, we combine them two-by-two with 3 random video samples to form 3 video-audio pairs, 15 in total.
The model takes these video-audio pairs as input and gets 3 generated samples for each reference audio, which will be used as test samples to compare with the reference audio.

For the ablation experiments, we only consider the reconstruction quality of the samples, so we randomly select 10 original pairs in the test set as input and obtain the generated samples.

\label{ap:mos}

\section{E. Ablation Studies}
\label{ap:abl}
\begin{table}[ht]
    \centering

    \begin{tabular}{c|c|c|c}
        \toprule
        \multirow{2}{*}{\textbf{Category}}&
        \multicolumn{3}{c}{\textbf{MCD Scores}}\\
        \cline{2-4}
        \multicolumn{1}{c|}{\textbf{}} &  
        \multicolumn{1}{c|}{\textbf{Proposed}} &
        \multicolumn{1}{c|}{\textbf{w/o MD}}& 
        \multicolumn{1}{c}{\textbf{w/o TDAD}}\\
        \midrule
        Baby & $4.77\pm 0.10)$ & $5.29(\pm 0.12)$  & $5.12(\pm 0.12)$ \\
        \midrule
        Cough &  $3.77(\pm 0.23)$ & $4.41(\pm 0.17)$ & $4.79(\pm 0.19)$  \\
        \midrule
        Dog & $4.19(\pm 0.18)$ & $4.22(\pm 0.20)$ & $4.44(\pm 0.14)$  \\
        \midrule
        Drum & $4.01(\pm 0.18)$ & $4.11(\pm 0.21)$ & $4.38(\pm 0.26)$  \\
        \midrule
        Fireworks & $3.64(\pm 0.10)$ & $3.61(\pm 0.10)$ & $3.68(\pm 0.12)$  \\
        \midrule
        Gun &  $3.73(\pm 0.20)$ & $3.73(\pm 0.14)$ & $3.77(\pm 0.15)$  \\
        \midrule
        Hammer & $4.05(\pm 0.27)$ & $4.02(\pm 0.17)$ & $4.39(\pm 0.13)$  \\
        \midrule
        Sneeze & $4.34(\pm 0.13)$ & $4.58(\pm 0.22)$ & $4.97(\pm 0.20)$  \\
        \midrule
        \textbf{Average} & $\mathbf{4.14(\pm 0.07)}$ &$4.25(\pm 0.08)$  & $4.43(\pm 0.08)$ \\
        \bottomrule
    \end{tabular}
    \caption{MCD scores without one discriminator. ``MD" denotes the Multi-Window Mel Discriminator and ``TDAD" denotes the Temporal Domain Alignment Discriminator. ``w/o" denotes the model retrained without some discriminator.}
    \label{tab:disabl}
    \vspace{-2mm}
\end{table}

\subsection{Generator Encoders Ablation}
\begin{figure}[ht]
\begin{center}
\vspace{3mm}
\includegraphics[width=\linewidth]{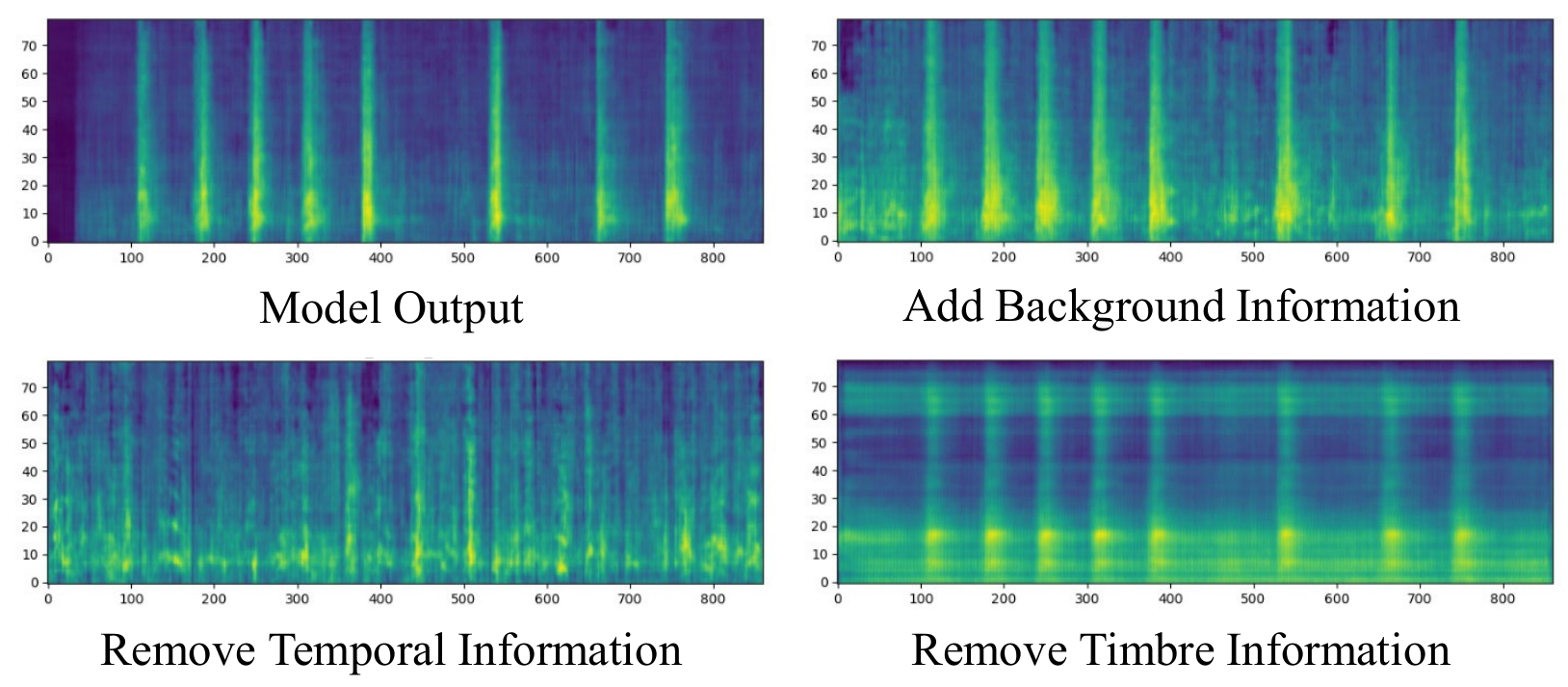}
\end{center}
\caption{Generated result when one component is removed or added.}
\label{fig:aber}
\end{figure}

\begin{table*}[h]
    \centering

    \begin{tabular}{c|c|c|c|c|c}
        \toprule
        \multirow{2}{*}{\textbf{Category}}&
        \multicolumn{5}{c}{\textbf{MCD Scores}}\\
        \cline{2-6}
        \multicolumn{1}{c|}{\textbf{}} &  
        \multicolumn{1}{c|}{\textbf{Proposed}} &
        \multicolumn{1}{c|}{\textbf{w Bi}} &
        \multicolumn{1}{c|}{\textbf{w/o Te}}& 
        \multicolumn{1}{c|}{\textbf{w/o Ti}} & 
        \multicolumn{1}{c}{\textbf{w/o Te \& Ti}} \\
        \midrule
        Baby & $4.77(\pm 0.10)$ & $3.75(\pm 0.25)$  & $5.26(\pm 0.16)$ & $6.84(\pm 0.26)$ & $6.60(\pm 0.29)$ \\
        \midrule
        Cough &  $3.77(\pm 0.23)$ & $3.72(\pm 0.14)$ & $3.87(\pm 0.23)$ & $5.16(\pm 0.20)$ & $5.15(\pm 0.24)$ \\
        \midrule
        Dog & $4.19(\pm 0.18)$ & $3.90(\pm 0.14)$ & $4.42(\pm 0.13)$ & $4.89(\pm 0.15)$ & $4.91(\pm 0.14)$ \\
        \midrule
        Drum & $4.01(\pm 0.18)$ & $3.38(\pm 0.16)$ & $4.10(\pm 0.18)$ & $5.36(\pm 0.30)$ & $5.39(\pm 0.30)$ \\
        \midrule
        Fireworks & $3.64(\pm 0.10)$ & $3.10(\pm 0.13)$ & $3.85(\pm 0.08)$ & $4.91(\pm 0.18)$ & $5.06(\pm 0.18)$ \\
        \midrule
        Gun &  $3.73(\pm 0.20)$ & $3.39(\pm 0.12)$ & $3.88(\pm 0.12)$ & $4.74(\pm 0.21)$ & $4.60(\pm 0.27)$ \\
        \midrule
        Hammer & $4.05(\pm 0.27)$ & $3.47(\pm 0.16)$ & $4.95(\pm 0.14)$ & $4.53(\pm 0.17)$ & $4.15(\pm 0.20)$ \\
        \midrule
        Sneeze & $4.34(\pm 0.13)$ & $4.04(\pm 0.18)$ & $4.50(\pm 0.15)$ & $5.91(\pm 0.38)$ & $5.96(\pm 0.35)$ \\
        \midrule
        \textbf{Average} & $\mathbf{4.14(\pm 0.07)}$ &$\mathbf{3.60(\pm 0.07)}$  & $4.33(\pm 0.08)$ & $5.25(\pm 0.12)$ & $5.21(\pm 0.12)$ \\
        \bottomrule
    \end{tabular}
    \caption{MCD scores with or without some components. ``Bi" denotes the Background Information,``Te" denotes the Temporal Information, ``Ti" denotes the Timbre Information. ``w" denotes the result generated with some components and ``w/o" denotes the result generated without some components.}
    \label{tab:abl}
\end{table*}
In the ablation experiment of the generator, we calculate the MCD scores for the generated result when one information component encoded by a certain encoder is removed as an objective evaluation.
As shown in Table \ref{tab:abl}, the generated results of our model achieve the second-lowest score on all experiments, while the results with the background information achieve the lowest score on all experiments.
The result is reasonable since there is a trade-off between audio reconstruction quality and audio quality due to the presence of background noise.
Specifically, the three parts of information are all necessary for the reconstruction of the target mel-spectrogram, and it will gain a larger distance between the generated result and the target audio when we discard one of the information, even if the information may conduct a negative impact on the quality of our generated results (e.g., the background noise).
Meanwhile, the above results can also corroborate the effectiveness of the background encoder in our model.
The results also show that timbre information has a more significant impact on the quality of the reconstructed audio than temporal information on average. 
The generated result of our model acquires the minimum MCD score, which has successfully demonstrated the effectiveness of the encoders of our model.
To better illustrate the above experimental results, we compare the reconstructed mel-spectrogram when one information component is removed or added (by setting the input vector to Gaussian noise or zero), and visualize the mel-spectrogram reconstruction results as shown in Figure \ref{fig:aber}.
As can be observed, when the temporal information is removed, the output mel-spectrogram becomes meaningless content that is uniformly distributed over the time series, and when the timbre information is removed, the output becomes a random timbre without specific spectral characteristics.
For the background information, when it is added during inference, the mel-spectrogram's background noise becomes brighter.

\subsection{Discriminators Ablation}

In the ablation experiments of discriminators, we retrain our model with one of the discriminators disabled. The experiments are performed under the same settings and configurations as before.  
As can be observed in Table \ref{tab:disabl}, the MCD scores of the generated results for almost all categories decreased to various extents after removing any of the discriminators.
On average, the impact of removing the Temporal Domain Alignment Discriminator is more significant than that of removing the Multi-window Mel Discriminator.
Due to the fact that the mel-spectrogram compresses the high-frequency components to some extent, some of the categories with high-frequency information content, such as \textit{Fireworks}, \textit{Gun}, and \textit{Hammer}, do not have significant differences in the scores obtained after removing the mel discriminator.

\section{F. Generated Result}
\label{ap:res}
\begin{figure*}[h]
    \centering
    \vspace{-2mm}
    \begin{minipage}[c]{\linewidth}
    \centering
    \includegraphics[width=\linewidth]{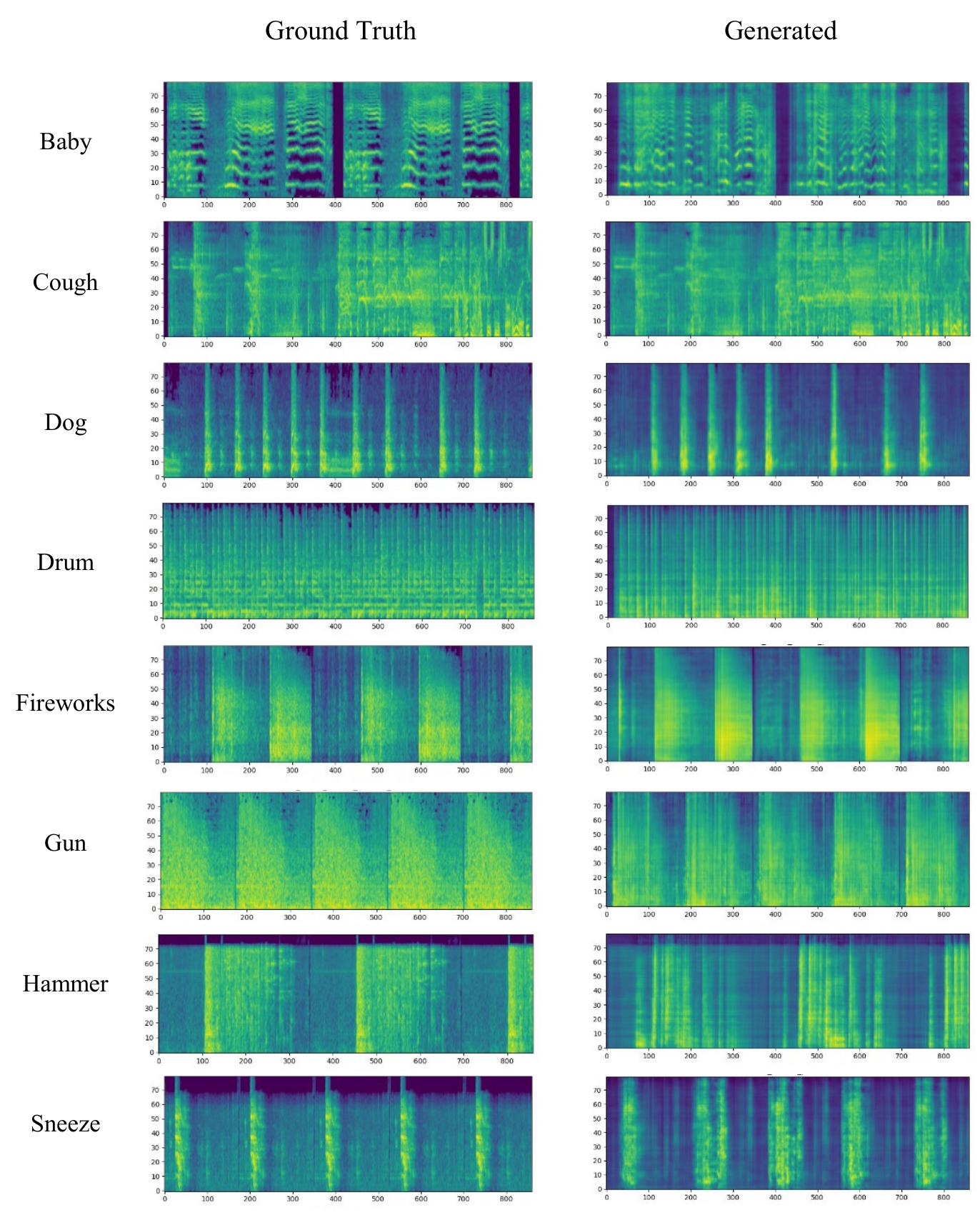}
    \end{minipage}%
    \vspace{-2mm}
    \caption{Mel-Spectrogram Comparison}
    \label{fig:mel}
    \vspace{-4mm}
\end{figure*}

As shown in Figure \ref{fig:mel} is a mel-spectrogram comparison of our generated result and the ground truth.
There may have some different between the two mel-spectrogram due to the out-screen sound and the background noise in the ground truth audio.
